\documentclass[aps,shownopacs,superscriptaddress,nofootinbib,preprint,11pt]{revtex4}
\usepackage{hyperref}
\usepackage{amsmath}
\usepackage{float}
\usepackage{graphics}
\usepackage{graphicx}
\usepackage{epsfig}
\usepackage{psfrag}
\usepackage{color}
\usepackage{slashed}

\usepackage{amsfonts}
\usepackage{amssymb}

\newcommand*{\be}{\begin{equation}}
\newcommand*{\ee}{\end{equation}}
\newcommand*{\bea}{\begin{eqnarray}}
\newcommand*{\eea}{\end{eqnarray}}

\newcommand{\comment}[1]{}

\newcommand{\cref}[1]{Chapter~\ref{c.#1}}

\def\beq{\begin{equation}}
\def\eeq{\end{equation}}
\def\bea{\begin{eqnarray}}
\def\eea{\end{eqnarray}}
\def\ba{\begin{array}}
\def\ea{\end{array}}
\def\bi{\begin{itemize}}
\def\ei{\end{itemize}}
\def\be{\begin{enumerate}}
\def\ee{\end{enumerate}}
\def\bc{\begin{center}}
\def\ec{\end{center}}
\def\bt{\begin{table}}
\def\et{\end{table}}
\def\btb{\begin{tabular}}
\def\etb{\end{tabular}}

\def\lsim{\raise0.3ex\hbox{$\;<$\kern-0.75em\raise-1.1ex\hbox{$\sim\;$}}}
\def\gsim{\raise0.3ex\hbox{$\;>$\kern-0.75em\raise-1.1ex\hbox{$\sim\;$}}}

\begin{document}

\title{Bulk Majorana mass terms and Dirac neutrinos in Randall Sundrum Model}
\author{Abhishek M Iyer}
\email{abhishek@cts.iisc.ernet.in}
\affiliation{Centre for High Energy Physics, Indian Institute of Science,
Bangalore 560012}
\author{ Sudhir K  Vempati}
\email{vempati@cts.iisc.ernet.in}
\affiliation{Centre for High Energy Physics, Indian Institute of Science,
Bangalore 560012}

\begin{abstract}
We present a novel scheme where Dirac neutrinos are realized even if lepton number violating Majorana mass terms 
are present. The setup is the Randall-Sundrum framework with bulk right handed neutrinos.  Bulk mass terms of both
 Majorana and Dirac type are considered. It is shown that  massless zero mode solutions exist when the bulk Dirac mass 
 term is set to zero. In this limit,  it  is found that the effective 4D small neutrino mass  is primarily of Dirac
 nature with the Majorana type contributions being negligible. Interestingly, this scenario is very similar to the one 
 known with flat extra dimensions. Neutrino phenomenology is discussed by fitting both
 charged lepton masses and neutrino masses simultaneously.  A single  Higgs localised on the IR brane is highly
 constrained as unnaturally large Yukawa couplings are required to fit charged lepton masses. A simple extension
 with two Higgs doublets is presented which facilitates a proper fit for the lepton masses. 
 
\end{abstract}
\vskip .5 true cm

\maketitle

\textbf{1.} As of today, we have no experimental indication whether neutrinos are of Majorana type or the Dirac type. 
On the theoretical side, models of both Dirac and Majorana type have been considered.  The popular 
seesaw mechanism with lepton number violation in the right handed neutrino sector \cite{Mohapatra} 
leads to small Majorana type masses for the left handed neutrinos.  
 Dirac type neutrinos on the other hand traditionally require lepton number conservation.  In most
models either this conservation is imposed by hand/construction or is a
 residue of some larger flavour symmetry \cite{10,11,12,13,gross,15,Dienes,17,19,20,21,22,23,Memenga}. Conservation
of global quantum numbers like lepton number is typically disfavored theoretically due to arguments based
on quantum gravity and worm holes \cite{Witten}.  For this reason, Dirac neutrinos are considered
to be some what unnatural.  

One possibility could be that lepton number is violated only by Planck scale operators. If these operators are then
some how suppressed, this would naturally pave way for Dirac neutrino masses 
\footnote{It should be noted that an alternate approach would be to consider discrete flavour symmetries which are imposed
to avoid Majorana mass terms and thus leading to Dirac neutrino masses \cite{Aranda:2013gga}.}.
In four dimensions the impact of the lepton
number violation at the Planck scale is characterized by the effective operator $LH.LH /M_{Pl}$ at the weak scale.
This leads to corrections to the neutrino mass matrix  $\sim ~\mathcal{O}(10^{-3}) ~\text{eV}$
if one assumes $\mathcal{O}(1)$ coefficients.
In higher dimensions explicit constructions can be done with specific Planck scale lepton number
 violating operators and their impact on weak scale physics can be studied.
 In fact in a particular example presented in \cite{planckgher} it has been shown that lepton number
 violation at the Planck scale can almost be hidden from weak scale physics.
 
In this case, a Randall-Sundrum (RS) \cite{RS} setup with two branes located at the two orbifold fixed points is considered.
The two fixed points, located at the $y=0$ and $y=\pi R$ are identified with the $UV\sim M_{Pl}$ and the $IR\sim \text{TeV}$ scales respectively.
The line-element for the RS background is given as
\begin{equation}
ds^2 =G_{MN}dx^Mdx^N= e^{-2\sigma(y)}\eta_{\mu\nu}dx^{\mu}dx^{\nu} - dy^2
\end{equation}
where $\sigma(y)=k|y|$.
Fermions and gauge bosons
are allowed to propagate in the bulk while the Higgs is localized on the IR brane
It has been shown that in the limit when the right handed neutrinos are highly IR localized,
whereas the Majorana mass terms are localised on the Planck brane, lepton number violating 
effects in effective neutrino mass matrix in four dimensions are highly suppressed.  Neutrino
masses can be of Dirac type by localised operators on the IR brane  \cite{planckgher}. The
main idea here being that the geometric `\textit{separation}' of the fields and the lepton number
violating operators leads to suppressed effects of the latter in the effective neutrino mass matrix. 
In the present letter we present a novel way of realizing Dirac masses in the same RS setup.
We will consider dimensionful Majorana mass terms as lepton number violating operators.
In particular we show that the lepton number violating operators need not be localized on the UV brane 
but instead can be present in the bulk. 
The magnitude of the operators can be as large as the Planck scale.
The 4D neutrinos are almost Dirac like,  in the limit when the 
bulk Dirac mass terms for the right handed neutrinos are set to zero. 
 Interestingly this case is very similar to the case discussed in flat extradimensions.

\textbf{2.} Extra space dimensions offer a new way of looking at the fermion mass hierarchy in the SM. 
Fermion bulk wave-functions are `split' and are localized at different points in the extra-dimension \cite{ArkaniHamed}. 
The point of localization is determined by the bulk Dirac mass parameters introduced separately for the left and right 
components of the 5D matter fields. The overlap of the zero mode wave-functions with the Higgs field determines the effective 
four dimensional Yukawa coupling of the fermion. In the Randall-Sundrum setup, where the bulk geometry is warped, 
localization of the fermions is natural; the point of localization is again determined by the bulk Dirac mass terms.
Neutrinos are however different from other matter fields as they allow both Dirac and Majorana mass terms.  
The profile of the zero mode crucially depends on the interplay between these two terms and the boundary conditions
one chooses.

We now give a brief review of bulk (right handed) neutrino fields in the RS set up. This case has been discussed in several 
papers \cite{warpedseesaw,a,b,c,d,planckgher,perez,Watanabe,iyer}. The leptonic and the quark part of the action is given by 

\begin{eqnarray}
\label{majoranab}
 S_{N}&=& S_{Kinetic}+\int d^4x\int dy \sqrt{-g}\left[ \frac{1}{2}\left(m_M\bar NN^c+\text{h.c.}\right)+ \ldots 
  \right. \nonumber \\ 
 &+& \left. \left( Y_N\bar L\tilde H N  + Y_E \bar{L} E H + \ldots  \right) ~\delta(y-\pi R)\right] \nonumber\\
 S_{Kinetic} &=& \int d^4x\int dy ~\sqrt{-g}~ \left(~ \bar{Q} (i\slashed D - m_Q)Q+\bar{u} (i\slashed D - m_u)u+\bar{d} (i\slashed D - m_d)d\right.\nonumber\\
&+& \left.\bar{N} (i\slashed D - m_N)N+\bar{L} (i\slashed D - m_{L})L+\bar{E} (i\slashed D - m_{E})E  ~\right).
 \end{eqnarray} 
 where the covariant derivative is defined ass
 \begin{equation}
\label{covariant}
 D_M =  \partial_M + \Omega_M +\frac{ig_5}{2}\tau^aW_{M}^a(x,y) + \frac{ig'}{2}Q_YB_M(x,y) 
\end{equation}
with $\Omega_M =  ( -k/2 e^{- k y} \gamma_\mu \gamma^5, 0)$ being the spin connection and $Q_Y$ is the hypercharge. $M$ is the five dimensional Lorentz index.
In the above,  $N$ ( $E$ ) are the neutrino (charged lepton ) singlet fields, $L$ are the lepton doublet fields 
and the Higgs field, is denoted by $H$ with $\tilde H= i\sigma_2H^*$. Generation indices have been suppressed. 
The bulk mass parameters for the $N$ fields are $m_M$ ($m_N$) for the Majorana (Dirac) type.  Here
$N^c = C_5 \bar{N}^{T}$ with $C_5$ being the five-dimensional charge conjugation matrix\footnote{$C_5$ is taken to be $C_4$.}.
$m_L$($m_E$) stands for the bulk (Dirac type) mass terms of the doublet (singlet) fields. All the bulk mass parameters are expressed in terms
of the so called $c$ parameters, for example,  $m_M =c_M k$, with $k$  being the reduced Planck scale.  Similarly, $m_N$ = $c_N k$ etc\footnote{In the
following, we will consider all the mass parameters to be real.}. $Y_{N,E}$ are the Yukawa parameters with mass dimensions, $ [Y] =  -1$. 
Finally, let us note that  in the above action we assumed the Higgs field to be localized on the IR brane.

The bulk fields can be Kaluza-Klein (KK) expanded in terms of their four dimensional fields and profiles in the
fifth direction. For the discussion relevant here, we consider the expansions of $N$ and $L$ fields as: 
\begin{eqnarray}
 N(x,y)=\sum_{n=0}^{\infty} {e^{2\sigma(y)} \over \sqrt{\pi R}} \left(N^{(n)}_1(x)g_1^{(n)}(y)+N^{(n)}_2(x)g_2^{(n)}(y)\right)\nonumber \\
 L(x,y)=\sum_{n=0}^{\infty}{e^{2\sigma(y)} \over \sqrt{\pi R}}  \left(L^{(n)}_L(x)f_L^{(n)}(y)+L^{(n)}_R(x)f_R^{(n)}(y)\right)
 \end{eqnarray}
 where $N_1^{(n)}(x)$ and $N_2^{(n)}(x)$ are the two Weyl components of the neutrino singlet field with $g_1^{(n)}(y)$ and $g_2^{(n)}(y)$ representing
 their profiles in the $y$ direction\footnote{We will specify the $Z_2$ properties of these components separately for each 
 case we consider.}.  Similarly, for the $L$ field, $L_L^{(n)}(x)$ and $L_R^{(n)}(x)$ represent the Weyl components 
 along with  $f_L^{(n)}(y)$ and $f_R^{(n)}(y)$ represent the respective profiles.  The profiles can be derived from the action
 after  imposing the ortho-normality conditions. For the KK modes of the $N$ field, the profiles are the solutions of the
 following couple differential equations \cite{warpedseesaw}
\begin{eqnarray}
(\partial_y+m_N)g_1^{(n)}(y)=m_ne^\sigma g_2^{(n)}(y)-m_Mg_2^{(n)}(y) \nonumber \\
(-\partial_y+m_N)g_2^{(n)}(y)=m_ne^\sigma g_1^{(n)}(y)-m_Mg_1^{(n)}(y)
\label{majcoupled}
\end{eqnarray}
A \textit{crucial} point to note is that  zero mode solutions, $m_n=0$ in  the set of equations in Eq.(\ref{majcoupled}),
are not consistent with the boundary conditions at the orbifold fixed points\cite{warpedseesaw}. 
Solutions however do exist for higher modes and they can be obtained numerically. 
A detailed phenomenological analysis can be found in \cite{iyer}.

\textbf{3.} Let us now revisit the result of Ref.\cite{planckgher} where Dirac neutrinos are realized in the above RS setup
with Majorana operators.
Zero mode solutions for Eq.(\ref{majcoupled}) are however possible if the Majorana mass terms are localized on the UV or IR boundary.
For the case  where they are confined to the UV boundary, the bulk eigenvalue equations for the $N$ fields 
in Eq.(\ref{majcoupled}) simply reduce  to 

\begin{eqnarray}
(\partial_y+m_N)g_1^{(n)}(y)=m_ne^\sigma g_2^{(n)}(y)\nonumber \\
(-\partial_y+m_N)g_2^{(n)}(y)=m_ne^\sigma g_1^{(n)}(y)
\label{majcoupled1}
\end{eqnarray}
Analytical solutions can easily be derived  for $m_n=0$.  Lets  consider $N_1$ component to be even under the
$Z_2$ symmetry and $N_2$ component to be odd. 
The  profile of $N_1^{(0)}= N_1$ is given by  $g_1^{(0)} (y) = g_1(y) =\mathcal{N}_Ne^{-c_Nky}$, where as 
the $Z_2$ odd field, $N_2$,  has no zero modes. 
The normalization factor $\mathcal{N}_N$ is given as 
\begin{equation}
 \mathcal{N}_N=\sqrt{\frac{0.5-c_N}{\epsilon^{2c_N-1}-1}}
 \label{normalizationfactor}
\end{equation}
where $\epsilon=e^{-kR\pi}$. For a regular RS setup $kR\sim 11.4$, which implies $\epsilon \sim 10^{-16}$.
It should be noted that profiles of the zero modes of $L, E$ fields also carry the same form.
The zero mode of $N_1$ field picks up a Majorana mass  due to the localised term at the UV boundary given by: 
\begin{eqnarray}
  m_{N^{(0)}}&\sim& m_M  g_1(0)^2
 \end{eqnarray}
Assuming $c_N<0.5$ and $m_M \sim k $, the above equation becomes
\begin{eqnarray}
 m_{N^{(0)}}&\sim & k ~(0.5-c_N)e^{-(1-2c_N)kR\pi}\nonumber \\
&\sim& 1 ~\text{TeV}~\epsilon^{-2c_N}
 \label{branemajorana}
 \end{eqnarray}
where $k\epsilon  \sim 1~ \text{TeV}$. 
 To analyze the neutrino mass matrix,  we should also consider the IR brane localised terms, the second line of  Eq.(\ref{majoranab}),
 generated from the Yukawa interaction. These are Dirac mass terms  and are given by 
\begin{equation}
m_{D_\nu}^{(0,0)} =\frac{v}{\sqrt{2}}g_1(\pi R) Y_N'f_L(\pi R)
\end{equation}
 where the $\mathcal{O}$(1) parameter $Y_N'=2kY_N$ and $f_L^{(0)}=\mathcal{N}_Le^{-c_Lky}$ denotes the zero mode profile of the doublet. The resultant neutrino mass matrix (with one KK mode ) has Type-I seesaw structure. 
In the basis $\eta^T = \{{ \nu_L^{(0)}, N_1^{(0)}}\}$, the Majorana
mass to the lowest order is given as
\begin{equation} 
 \mathcal{L}_m = - {1 \over 2} \eta^T \mathcal{M}_\nu \eta \;\;\;; \;\;\;\;\mathcal{M}_\nu=\begin{pmatrix}
      0&m_{D_\nu} \\
      m_{D_\nu}&m_{N^{(0)}}
      \end{pmatrix}
\label{majoranamatrix}
      \end{equation}
where we have  assumed one flavour for simplicity.
From Eq.(\ref{branemajorana}) we see that as
$c_N\rightarrow -\infty$, $m_{N^{(0)}}\rightarrow 0$ \footnote{Note that $c_N\sim -1$ is sufficient to make $m_{N^{(0)}}$ insignificant.}.
Note that this limit holds while $c_M$ is taken to be $\mathcal{O}$(1) and the 
Majorana mass terms can be $\mathcal{O}(M_{PL})$.
As a result the Majorana mass for the right handed neutrino almost vanishes.  In this limit, the eigenvalues of the neutrino mass matrix in Eq.(\ref{majoranamatrix})
are $\pm m_{D_\nu}^{(0,0)}$. This implies that the localization of the zero mode of the neutrino singlet very close to the IR brane results in its negligible 
overlap with the lepton number violating operator situated on the UV brane.
The small neutrino masses are determined entirely by the brane localized Yukawa coupling in Eq.(\ref{majoranab}) thus attributing
a Dirac nature to the neutrinos.

\textbf{4.} We now discuss the alternative possibility of realizing Dirac neutrinos in the presence of lepton number violating
terms. However we will assume $m_N=0$ in Eq.(\ref{majoranab}). Bulk flavour symmetry groups can be imposed to achieve this limit.
For example consider the following bulk flavour symmetry group for leptons
\begin{equation}
 G_{lepton}\equiv SU(3)_L\times SU(3)_E\times O(3)_{N_R}
 \label{flavourgroup}
\end{equation}
The transformation of the leptonic fields under $G_{lepton}$ is given as 
\begin{eqnarray}
 L\sim (3,1,1)\;\;\;E\sim (1,3,1)\;\;\;\ N_{R}\sim (1,1,3)\;\;\;\; N_{L_i}\sim(1,1,1)~ (i=1,2,3)
\end{eqnarray}
Note that we have given different representations for the left and right chiralities of the $N$ field. The $Z_2$ odd field $N_L$ is considered to transform as a singlet under $O(3)$.
This choice leads to a vanishing bulk Dirac mass ($m_N=0$) for the singlet field.
In this case, the Majorana mass terms are no longer localised on the UV brane, but are present in the bulk. 
In this limit where the bulk Dirac mass for $N$ vanishes, the solutions to Eq.(\ref{majcoupled}) are simple  
to obtain and are given as
\begin{eqnarray}
g_1^{(n)}(y) &=& \xi\sin(\frac{m_ne^\sigma}{k}-m_My), \nonumber \\
g_2^{(n)} (y) &=&\xi \cos(\frac{m_ne^\sigma}{k}-m_My ) 
\label{wavefunction}
\end{eqnarray}
where $\xi\sim\sqrt{\pi Rk}e^{-0.5\sigma(\pi R)}$.
Imposing proper boundary conditions we can project out the $Z_2$ odd components on the boundary. 
For example, we choose as before that the $N_2$ component is $Z_2$ odd and $N_1$ is the $Z_2$ 
even component. The boundary conditions for which the $Z_2$ odd part say $g_2$ vanishes on the boundary 
are given as
\begin{equation}
\frac{m_ne^\sigma}{k}-m_M\pi R=(2n+1){ \pi \over 2}
\label{boundaryconditions}
\end{equation}
where n=0,1,2\ldots. The zero mode\footnote{In principle massless modes are also possible by choosing $m_M=\frac{-p}{R}$ where $p\in \mathcal{Z}^+$.}($n=0$, massless solutions) can exist if $m_M$ takes values 
 $m_M=\frac{-1}{2 R}$. 
 In this case, we have 
\begin{eqnarray}
g_1^{(0)} (y)&=&\xi\sin(-m_My )\nonumber \\ 
g_2^{(0)} (y)&=&\xi\cos(m_My )
\end{eqnarray}
This corresponds to $c_M=\frac{-1}{2kR}$.  
The masses of the higher KK modes are determined from the boundary conditions in Eq.(\ref{boundaryconditions}) and are 
given as
\begin{equation}
 m_n \sim nk\pi\epsilon\;\;\; ;\;n=1,2,3\ldots
\end{equation}
Unlike the other RS fields, the KK modes for the $N$ field are regularly spaced at $1\pi$ TeV, $2\pi$ TeV etc. This reminds us
of the KK bulk fields in Arkani-Hamed, Dimopouplos, Dvali (ADD) models \cite{ADD,Antoniadis:1998ig}.  Thus if one considers bulk Majorana mass terms
instead of Dirac mass terms,  the profiles in the bulk for the zero and higher modes  are periodic.

Let us consider the total neutrino mass matrix in this case.  In the basis, $\chi^T = \{ \nu_L^{(0)}, N^{(0)}_1, N^{(1)}_1\ldots\}$ 
the neutrino mass matrix takes the form 
\begin{equation}
\mathcal{L}_m = - {1 \over 2} \chi^T \mathcal{M} \chi \;\;\;; \;\;\;\; \mathcal{M} =   \begin{pmatrix} 
0 & m_{D_\nu}^{(0,0)} & -m_{D_\nu}^{(0,1)}&m_{D_\nu}^{(0,1)}&\ldots\\
m_{D_\nu}^{(0,0)} &0& 0&0&\ldots \\
-m_{D_\nu}^{(0,1)} & 0 & m_{1}&0&\ldots\\
m_{D_\nu}^{(0,1)}&0&0&m_{2}&\ldots\\
\vdots&\vdots&\vdots&\vdots&\ddots\\
\end{pmatrix} 
\label{majoranamass}
\end{equation}
where the 4D Dirac mass for the neutrino induced on the IR brane is given as 
\begin{equation}
 m_{D_\nu}^{(0,0)}=\frac{v}{\sqrt{2}}Y'_N\sqrt{\frac{0.5-c_L}{\epsilon^{2c_L-1}-1}}\epsilon^{c_L-0.5}
\label{neutrinomass}
 \end{equation}
where $Y'_N=2kY_N$ and $|m_{D_\nu}^{(0,0)}|=|m_{D_\nu}^{(0,i)}|$ $\forall$ $i\geq1$.  The form of this 
mass matrix is very similar to the one  with bulk right singlet neutrinos and brane
localised left handed doublet lepton fields in flat extra dimensions considered in Ref.\cite{Dienes}. 
The analysis of Ref. \cite{Dienes}
can be used to find the eigenvalues of the mass matrix (\ref{majoranamass}). 

If there are $n_0$ KK modes in the theory, the effective neutrino mass matrix in the basis $(\nu_L^{(0)}, N^{(0)}_1)$, 
after integrating our the KK modes is given as \cite{Dienes} 
\begin{equation}
\mathcal{M}_{eff}=\begin{pmatrix}
 a_0&m_{D_\nu}^{(0,0)}\\
 m_{D_\nu}^{(0,0)}&0
\end{pmatrix}
\end{equation}
where $a_0\sim -\frac{m^2R}{\epsilon}\ln{n_0}$.
The neutrino mass eigenvalues can be written as
\begin{equation}
 m_\nu\sim\pm m_{D_\nu}^{(0,0)} -\frac{m^2R}{\epsilon}\ln{n_0}
 \label{addegs}
\end{equation}

In the limit where $a_0 ~\ll~ m_{D_\nu^{(0,0)}}$, neutrinos are automatically Dirac-like.
The second term in Eq.(\ref{addegs}) is the Majorana ``seesaw" like terms which is a result of integrating out
heavy KK modes.
In contrast with the ADD case, this
limit is natural in the RS case. In the RS case, $R$ is small, $\sim {1 \over k}$, which makes the contribution from the Majorana
term negligible. This result holds even for a large $n_0 \sim 10^{18}$.  On the other hand, in the ADD case, a very large
radius,  $R \sim \text{eV}^{-1}$ is required to have a large contribution  from the ``seesaw" term in the limit $n_0$ 
becomes very large.  $m_{D_{\nu}}^{(0,0)}$ can  be made small $\sim m_{atm}$ by choosing $c_L$ values appropriately. 
For example, this can be achieved by localizing the leptonic doublets close to the UV brane.  

Finally let us note that the above situation
can be easily generalized to three generations with three bulk right handed neutrinos. In the next section we discuss
neutrino phenomenology in detail.

\textbf{5.} 
We simultaneously fit neutrino mass differences and charged lepton masses along with the mixing angles to the $c$ parameters
and the $\mathcal{O}(1)$ Yukawa couplings. More details of the fit can be found in \cite{iyer}.  For the standard RS set
up, $\epsilon \sim 10^{-16}$, we find that to reproduce the atmosphere neutrino mass scale,  $\mathcal{O}$(0.03) eV we need
a $c_L\sim 1.3$. 
We have chosen the $\mathcal{O}(1)$ Yukawa coupling $Y'$ to be 1. Such $c$  values for the
lepton doublets make it difficult to fit simultaneously the charged lepton masses\footnote{Such a situation was also encountered
in the case neutrinos get their masses through higher dimensional  lepton number violating operator \cite{iyer}.}.  
The charged lepton mass matrix is given as
\begin{equation}
 m_{ij}=(Y'_E)_{ij}\mathcal{N}_{L_i}\mathcal{N}_{E_j}\epsilon^{c_{L_i}+c_{E_j}-1}
\label{c1}
 \end{equation}
where $\mathcal{N}_{L,E}$ have the same form as Eq.(\ref{normalizationfactor}) and the dimensionless 
$\mathcal{O}(1)$ Yukawa coupling for a brane localized Higgs is defined as $Y'_E=2kY_E$.
The required
bulk mass parameters for the charged lepton fields, $c_E$ turn out to be large and negative. This introduces a host of other
problems like non-perturbative Yukawa couplings etc. On the other hand, changing the warp factor will not have much impact
on the results.  For example, for a $\epsilon \sim 10^{-2}$, we find that $c_L$ values required are even larger.  


A simple solution would be to disentangle the Higgs fields responsible for neutrino masses and charged leptons by introducing
an additional Higgs doublet as in a two Higgs doublet model. We denote the Higgs doublet generating the Dirac neutrino masses
by $H_u$ and the other as $H_d$. $H_u$ is localised on the  IR brane where as the  $H_d$ is a bulk Higgs field, whose localisation 
is fixed by the charged lepton masses. The Yukawa part of the lagrangian is now  given as

\begin{equation}
 L_{Yuk}\subset \int d^4x dy \left[\left(\delta(y-\pi R)Y_N\bar LH_uN+Y_E\bar LH_dE\ldots\right)\right]
\end{equation}
As before, the neutrino masses are given by Eq.(\ref{neutrinomass}) with $H$ replaced by $H_u$.
 To determine  the charged lepton mass matrices with a bulk Higgs, we briefly review the 
the derivation for the profile equation as well as the Yukawa couplings for a bulk Higgs.

For a bulk scalar field in a warped background, the presence of zero modes requires the addition 
of brane localized mass terms.  For the bulk field $H_d$, the  action is given as \cite{gherghetta,Gherghetta1} 
\begin{equation}
 S=\int d^4xdy\sqrt{-g}\left[\partial_MH_d^*\partial^MH_d+\left[m_{H_d}^2+2bk\left(\delta(y)-\delta(y-\pi R)\right)\right]|H_d|^2\right]
\end{equation}
where we parametrize the bulk mass as $m_{H_d}^2=ak^2$ with $a,b$ being dimensionless quantities. Ideally one would expect them to be $\mathcal{O}$(1).
The zero mode profile for a bulk scalar is given as 
\begin{equation}
 f_{H_d^{(0)}}=\sqrt{k\pi R}\zeta_{H_d} e^{(b-1)ky}
\end{equation}
where the normalization factor $\zeta_\phi$ is given as
\begin{equation}
\zeta_{H_d}=\sqrt{\frac{2(b-1)}{\epsilon^{2(1-b)}-1}}
\end{equation}
The brane parameter $b$ must be tuned to be $b=2\pm\sqrt{4+a}$ to satisfy the boundary conditions for the zero modes. $b>1(b<1)$ implies the zero mode of the Higgs is localized towards the IR(UV) brane.

For a bulk Higgs the fundamental Yukawa couplings $Y_E$ have mass dimension -1/2. After performing the KK expansion and integrating
over the extra-dimension the zero mode mass matrix for all charged leptons in general is given as \footnote{The down type quarks have the same form of the mass matrix
as the charged leptons while
the up type quark mass matrix is similar to Eq.(\ref{c1}).}

\begin{equation}
 m_{ij}=v_d(Y'_E)_{ij}\zeta_{H_d} \mathcal{N}_{{L_i}} \mathcal{N}_{{E_j}}\left(\frac{\epsilon^{(c_{L_i}+c_{E_j}-b)} -1}{b-c_{L_i}-c_{E_j}}\right)
 \label{diracmass}
\end{equation}
where we have defined the dimensionless $\mathcal{O}$(1) Yukawa coupling as $Y'_E=2\sqrt{k}Y_E$ and the normalization factor $\mathcal{N}_i$ is defined in Eq.(\ref{normalizationfactor}).
The corresponding $\mathcal{O}$(1) parameters for the up and the down sector quarks are denoted as $Y'_U$ and $Y'_d$ respectively. 

The ratio of the vev of the two Higgs doublet is defined as $tan\beta=\frac{v_u}{v_d}$.
We choose $tan\beta=10$ for illustration.
While $H_u$ is localized on the IR brane, $H_d$ is localized near the UV brane ($b<1$). 
 Fitting Eq.(\ref{neutrinomass}) for small neutrino masses require $c$ value for the doublet to be $\sim 1.3$.
 Corresponding to this and fitting Eq.(\ref{diracmass}) for the charged leptons we choose $b=0.3$. We find that the electron mass can be conveniently fit by choosing $c_{E_R}\sim 0.3$ 
 while the remaining charged leptons can be fit by choosing
 a range $0.4<c<1$ for the corresponding bulk mass parameters of the singlets. Table[\ref{fit1}] shows the range of $c$  parameters obtained which fit the lepton
masses and mixing angles. For the leptonic case we assume normal hierarchy of neutrino mass eigenvalues. The magnitude of $\mathcal{O}(1)$ Yukawa parameters in the
leptonic sector \textit{i.e.} $Y'_{N,E}$ were chosen to be in the range $[0.08,4]$ to fit the data.
 This configuration of Higges can also accommodate quark masses by the introduction of 9 bulk mass parameters \textit{i.e} $c_Q,c_U,c_d$ in Eq.(\ref{majoranab}).
 To fit the top quark mass, the third generation top singlet require $c_{t_R}\leq -3.0$ while the lighter generations including the third 
generation doublet can be fit by choosing the corresponding $c$ values to be $0.3<c<1$. Table[\ref{fit2}] shows the range of $c$  parameters for the hadronic sector
which fit the quark masses and CKM angles. 
The magnitude of $\mathcal{O}$(1) Yukawa parameters for the quark sector were also chosen to lie between $0.08<|Y'_a|<4$ where $a=d,U$. 

\begin{table}[htdp]
\begin{center} 
\begin{tabular}{|cccc|}

\hline
parameter & range&parameter & range\\

\hline
$c_{L_1}$ & 1.27-2.4 &$c_{E_1}$ &0.19-0.36 \\
$c_{L_2}$ & 1.26-1.4 &$c_{E_2} $&0.35-0.40 \\
$c_{L_3}$ & 1.25-1.38 &$c_{E_3} $&0.44-0.52 \\
\hline 
\end{tabular}
\end{center}
\caption{Range of $c$ parameters which fit the lepton masses and mixing angles in the bulk two Higgs doublet model with bulk Majorana masses.
$\mathcal{O}$(1) Yukawa parameters are chosen to lie between $0.08<|Y'|<4$. Normal hierarchy is assumed for the neutrino masses.}
\label{fit1}
\end{table}

\begin{table}[htdp]
\begin{center} 
\begin{tabular}{|cccccc|}
\hline

parameter & range &parameter & range&parameter & range\\

\hline
$c_{Q_1}$&0.39-0.57&$c_{D_1}$&0.30-0.49&$c_{U_1}$&0.67-0.79 \\
$c_{Q_2}$&0.47-0.55&$c_{D_2}$&0.37-0.93&$c_{U_2}$&0.53-0.58 \\
$c_{Q_3}$&0.502-0.507&$c_{D_3}$&0.68-0.97&$c_{U_3}$& $-6.8$ - $-3.0$ \\
\hline 
\end{tabular}
\end{center}
\caption{Range of $c$ parameters which fit the quark masses and mixing angles in the bulk two Higgs doublet model with bulk Majorana masses.
$\mathcal{O}$(1) Yukawa parameters are chosen to lie between $0.08<|Y'|<4$. }
\label{fit2}
\end{table}

\textbf{6.} 
Nature has not yet spoken whether neutrinos are Dirac or Majorana. While theoretically we are prejudiced to consider that
Majorana neutrinos are more natural as quantum gravity does not conserve global symmetries, it is not uncommon to find
examples where Dirac neutrinos can exist even with lepton number violation.  
In the present work, using the RS set up, we presented a scenario in which  Dirac neutrinos can be obtained in the presence 
of lepton number violating terms in the bulk.  

We have considered lepton localisation purely with bulk Majorana mass terms. Bulk Dirac mass terms are set to zero. This case
leads to periodic KK modes similar to ADD models. Zero modes can exist for particular values of the Majorana mass terms. 
In this case, the Majorana contribution can be shown to be negligible leading to Dirac neutrinos.  Phenomenologically, 
this model requires large $c_E$ values for the bulk mass parameters which is problematic to fit charged lepton masses. 
A simple extension in terms of two Higgs doublet model is presented where a good fit to neutrino masses and charged lepton
masses is simultaneously obtained. We have not commented about electroweak precision constraints nor flavour constraints
in this model as our focus has been purely on fermion masses. Electroweak precision constraints are expected to be strong
and one possible way out is to consider a much larger gauge group and particle spectrum towards a custodially symmetric RS model \cite{Agashe:2003zs}.
We expect that inclusion of both the Higgses should be straightforward. On the other hand flavour is expected to be strong
due to presence of new Higgs contributions. Fortunately, one can utilise Minimal Flavour Violation (MFV) techniques
to reduce flavour violation. For the flavour symmetry group presented in Eq.(\ref{flavourgroup}), the Yukawa couplings transform as
\begin{equation}
 Y_E\sim(3,1,\bar 3)\;\;\;\;\; Y_N\sim(3,1,3)
\end{equation}
The bulk mass parameters can be expressed in terms of the Yukawa as \cite{Fitzpatrick}
\begin{equation}
 c_L=I+\alpha Y_EY_E^\dagger+\alpha' Y_NY_N^\dagger\;\;\;\;\; c_E=\beta Y_E^\dagger Y_E\;\;\;\;\; c_N=0
\end{equation}
where $\alpha,\alpha',\beta\in\mathcal{R}$.
This leads to a strong suppression of flavour violation. A detailed analysis of flavour violation in RS models with two
Higgs doublets will be presented elsewhere \cite{ourrs2}.
The investigation of the fermion mass hierarchy in two Higgs doublets models and the resulting implications is interesting 
enough to be considered in greater detail.

\vskip 1 cm
 \noindent
\textbf{Acknowledgement}\\

SKV acknowledges support from DST Ramanujam fellowship  SR/S2/RJN-25/2008 of Govt. of India.

\bibliographystyle{ieeetr}

       \bibliography{majorana.bib}

\end{document}